\documentstyle[preprint,prl,aps]{revtex}
\renewcommand{\baselinestretch}{2}

\begin{document}
\draft

\title{Systematic derivation of a rotationally covariant extension of the
2-dimensional Newell-Whitehead-Segel equation}

\author{Robert Graham\\
Fachbereich Physik, Universit\"at-Gesamthochschule Essen\\
45117 Essen\\Germany}

\maketitle

\begin{abstract}
An extension of the Newell-Whitehead-Segel amplitude equation covariant
under abritrary rotations is derived systematically by the renormalization
group method.
\end{abstract}

\narrowtext

The present day theory of pattern formation to a large degree rests on
the derivation and exploitation of amplitude equations. In particular, for
the formation of 2-dimensional patterns by spontaneous symmetry breaking
at a finite wavenumber in isotropic 2-dimensional layers the amplitude
equation due to Newell and Whitehead and to Segel \cite{1} is of
fundamental importance. It is derived, e.g. for the Benard instability,
close to the onset of the pattern, by assuming the presence of an ideal
one-dimensional roll pattern and certain scaling properties of its small
amplitude and its spatial and temporal variations.
The small parameter is the difference of the bifurcation parameter from
its value at onset.

The resulting amplitude equation, despite of
its great fundamental and practical importance, has a well known
short-coming: it does not respect the full rotation invariance of the
two-dimensional system and the fact, that in principle the rotational
symmetry is broken by the pattern {\it spontaneously} rather than by
any external agent. This short-coming is a direct consequence of the
method of derivation, which singles out a particular direction for
the main pattern, and even more importantly, makes use of
anisotropic assumptions for the scaling of the spatial variation of
the pattern.

Two recent developments now permit to overcome this problem \cite{1a}.
First, Gunaratne et al\cite{2} recently proposed an extended form
of the Newell-Whitehead-Segel (N-W-S) equation which they demonstrated to
fully respect the rotational symmetry of the original system. They also
provided a derivation of their equation by including, in a given order
in the bifurcation parameter, symmetry restoring terms, which,
indeed appear in higher order of the N-W-S scheme. While this
recombining of terms from different orders of the expansion
suggests that their equation is the correct symmetric extension of the
N-W-S equation, before one can accept this as a fact, a fully systematic
(if not mathematically rigorous) derivation of this equation must
be provided. Clearly, a new method, avoiding any anisotropic scaling
assumptions, is required. Recently, such a method has become available
with the renormalization group method by Chen et al \cite{3,4},
applied to the derivation of amplitude equations \cite{4,5}. Here
we shall employ this method to give a systemtic derivation of the
rotationally symmetric extension of the N-W-S equation. Curiously, Chen
et al \cite{4} and also Kunihiro \cite{5} already  used this method
to rederive the N-W-S equation, seemingly without invoking the anisotropic
scaling assumption of the original derivation. However, as we shall see,
such scaling assumptions in fact crept inadvertently into their derivation
by the assignement which out of a number of space-time secular terms
would be `most singular'. Avoiding the scaling assumption implicit in
their choice and simply keeping {\it all} singular terms, but otherwise
following closely the treatment in \cite{4} we are able to achieve our
goal.

For concreteness and simplicity, we shall present the derivation starting
from the two-dimensional Swift-Hohenberg equation \cite{6}
\begin{equation}
\frac{\partial\psi}{\partial t}=\epsilon(\psi-\psi^3)-
\left(\frac{\partial^2}{\partial x^2}+\frac{\partial^2}{\partial y^2}+
k^2\right)^2\psi
\label{eq:1}
\end{equation}
where $\epsilon$ is the bifurcation parameter, and $\psi$ was scaled to
make the $\psi$ and $\psi^3$-terms both of the same order in $\epsilon$.

The derivation starts with the ansatz
\begin{equation}
\psi=2{\rm Re}\left\{Ae^{ikx}+\epsilon\psi_1(x,y)+\cdots\right\}
\label{eq:2}
\end{equation}
with amplitude $A$ which is constant to lowest order in $\epsilon$.
The ansatz (\ref{eq:2}) breaks the
rotational symmetry in the $(x,y)$-plane. However,
due to the rotationally symmetric form of eq.~(\ref{eq:1}) the resulting
amplitude equation for $A$ will actually be covariant under rotations of
$\bbox{k}=k\hat{e_x}$ in eq.~(\ref{eq:2}) in a sense to be made concrete
below. Inserting
eq.~(\ref{eq:2}) in eq.~(\ref{eq:1}) one obtains an equation
\begin{equation}
\left(\sum_{i=1}^7L_i\right)\psi_1=(1-3|A|^2)Ae^{ikx}+
\;\mbox{\rm non-resonant terms}
\label{eq:3}
\end{equation}
with
\begin{eqnarray*}
 L_1 &=& \frac{\partial}{\partial t},\, L_2=
    \left(\frac{\partial}{\partial x}-ik\right)^4,\,
     L_3=4ik\left(\frac{\partial}{\partial x}-ik\right)^3,\\
L_4 &=& -4k^2\left(\frac{\partial}{\partial x}-ik\right)^2,\,
     L_5=2\left(\frac{\partial}{\partial x}-ik\right)^2
       \frac{\partial^2}{\partial y^2}\\
L_6 &=& 4ik\left(\frac{\partial}{\partial x}-ik\right)
         \frac{\partial^2}{\partial y^2},\,
 L_7=\frac{\partial^4}{\partial y^4}\,.
\end{eqnarray*}
The $L_i$ turn out all to commute. Now, the space-time secular part of a
general solution of eq.~(\ref{eq:3}) is constructed in the form
\begin{eqnarray}
&& \psi=2{\rm Re}\bigg\{Ae^{ikx}
   +\epsilon(1-3|A|^2)
     Ae^{ikx}\bigg(C_1t+C_2\frac{x^4}{24}+
           C_3\frac{x^3}{24ik}+
           C_4\frac{-x^2}{8k^2}\nonumber\\
&&\mbox{\hspace{6.5cm}}+C_5\frac{x^2y^2}{8}+
           C_6\frac{xy^2}{8ik}+C_7\frac{y^4}{24}\bigg)+\cdots\bigg\}
\label{eq:4}
\end{eqnarray}
where we have displayed all secular terms to order $\epsilon$ and the
dots denote non-secular or higher order
terms. The $C_i$ are integration constants and satisfy $\sum_iC_i=1$.
Incidentally, in \cite{4} and \cite{5} the terms $\sim t,x^2$, $xy^2$,
$y^4$ were considered as `most singular' and only those terms were
kept there. Because this assignment implies different scalings for the
other terms and breaks the rotational symmetry we shall not use it
here. This accounts for the differences of our results below from those
in \cite{4,5}.

The renormalization group method \cite{4} now proceeds by (i) introducing
an arbitrary `regularization point' $X$, $Y$, $T$ and to reorganize
eq.~(\ref{eq:4}) as an expansion around this point, rather than the origin;
(ii) splitting in the amplitudes of the secular terms
$x^n=(x^n-X^n)+X^n$ and similar for $y^n$, $t^n$ where $n$ is any of the
exponents of $x$, $y$, $t$ appearing there; (iii) absorbing
$X^n$, $Y^n$, $T^n$, by a redefinition of the amplitude $A$ which thereby
becomes a function $A(X,Y,T)$. The equation we are left with is
eq.~(\ref{eq:4}) with the replacements $A\to A(X,Y,T)$,
$t\to t-T$, $x^n\to x^n-X^n$, $y^n\to y^n-Y^n$, but $e^{ikx}\to e^{ikx}$.

The symmetry which gives rise to the renormalization group is the fact that
 $\psi$ is independent of $X,Y,T$. Therefore, all derivatives of
$\psi$ with respect $X,Y,T$ must vanish. To first order in
$\epsilon$ one obtains the conditions
\begin{eqnarray*}
\frac{\partial A}{\partial T} &-&\epsilon C_1A(1-3|A|^2)=0\\
\frac{\partial^4A}{\partial X^4}&-&\epsilon C_2A(1-3|A|^2)=0\\
4ik\frac{\partial^3A}{\partial X^3}&-&\epsilon C_3A(1-3|A|^2)=0\\
-4k^2\frac{\partial^2A}{\partial X^2}
    &-&\epsilon C_4A(1-3|A|^2)=0\\
2\frac{\partial^4 A}{\partial X^2\partial Y^2}
    &-&\epsilon C_5A(1-3|A|^2)=0\\
4ik\frac{\partial^3 A}{\partial X\partial Y^2}
    &-&\epsilon C_6A(1-3|A|^2)=0\\
\frac{\partial^4A}{\partial Y^4}&-&\epsilon C_7A(1-3|A|^2)=0
 \end{eqnarray*}
Adding these equations by using $\sum_iC_i=1$ and identifying
$X,Y,T\to x,y,t$ one obtains $\psi=2 Re\{A(x,y,z,t)e^{ikx}\}$ instead
of (\ref{eq:4}) and the amplitude equation, independent of the
$C_i$, to order $\epsilon$
\begin{equation}
\frac{\partial A}{\partial t}-\epsilon A(1-3|A|^2)=4k^2\Box_x^2A
\label{eq:5}
\end{equation}
with the symbol $\Box_x$ introduced in \cite{2} as
\[
\Box_x=\bbox{\hat{e}}_x\cdot\nabla-\frac{i}{2k}\nabla^2\,.
\]
The N-W-S equation is also of the form (\ref{eq:5}), but there
$\Box_x^2$ is replaced by $(\frac{\partial}{\partial x}-\frac{i}{2k}
\frac{\partial^2}{\partial y^2})^2$. It is obtained for special solutions
of (\ref{eq:3}) with $C_2=C_3=C_5=0$.
This completes our derivation. It is clear that the method is systematic
and may be extended to higher order in $\epsilon$. It can be seen that
the particular form $\Box_x$ of the spatial derivative operator is a
direct consequence of the form of all secular terms in (\ref{eq:4}),
which are in turn a consequence of the rotational invariance of the
derivative operator in (\ref{eq:1}). Hence $\Box_x$ will automatically
be generated by this method starting from any rotationally invariant
set of equations in place of (\ref{eq:1}), like e.g. the Boussinesq
equations for thermal convection.

The covariance of eq.~(\ref{eq:5}) under rotations has been discussed
in \cite{2}. It rests on the remarkable property
$\Box_xe^{i\Delta\bbox{k}_x\bbox{x}}=0$ \cite{2}, where
$\Delta\bbox{k}_x=\bbox{k}(\vartheta)-k\hat{\bbox{e}}_x=
k((\cos\vartheta-1)
\hat{\bbox{e}}_x+\sin\vartheta\hat{\bbox{e}}_y)$,
and the even more remarkable property
\[e^{-i\Delta\bbox{k}_x\bbox{x}}
\Box_xe^{i\Delta\bbox{k}_x\bbox{x}}=\Box_{x'}\,,
\]
where $\hat{\bbox{e}}_{x'}=
\cos\vartheta\hat{\bbox{e}}_x+\sin\vartheta\hat{\bbox{e}}_y$. It is the
second property
(which is independent of the
first because $\Box_x$ is of second order) which makes in eq.~(\ref{eq:5})
the transformation of the amplitude $A\to Ae^{i\Delta\bbox{k}_x
\cdot\bbox{x}}$
equivalent to the rotation $k\hat{\bbox{e}}_x\to k\hat{\bbox{e}}_{x'}$
of the wave-vector.

As many properties of patterns like e.g. defects and textures depend in
a crucial manner
on the underlying symmetry and the fact that it is spontaneously, not
externally, broken, the covariance of eq.~(\ref{eq:5}) under rotations
must be considered an important improvement
of the fundamental N-W-S equation. In addition we may remark that the
derivation we have given would still
apply if noise is added on the right hand side of (\ref{eq:1}) providing
a rotationally covariant extension of the noisy amplitude equation
derived in \cite{7}.

Like the N-W-S equation the amplitude equation
(\ref{eq:5}) has a potential, i.e. it can be written in the form
$\frac{\partial A}{\partial t}=-\frac{\delta\phi}{\delta A^*}$ with the
Lyapunov functional
\begin{equation}
  \phi = \int_{dxdy}\left\{-\epsilon|A|^2\left(1-\frac{3}{2}|A|^2\right)
         +4k^2|\Box_xA|^2\right\}\,.
\label{eq:6}
\end{equation}
However, unlike the potential for the N-W-S equation \cite{7}
(where again $\Box_x$ is replaced by $(\frac{\partial}
{\partial x}-\frac{i}{2k}
\frac{\partial^2}{\partial y^2})$) the potential (\ref{eq:6}) has minima,
the attractors of eq.~(\ref{eq:5}), for $A=(3)^{-1/2}
e^{i\Delta\bf{k}(\vartheta)\cdot\bf{x}}$ with {\it arbitrary}
angle $\vartheta$. In the noisy case W$(\{A,A^*\})=$ $N\exp[-\phi/Q]$
is the steady-state distribution of the fluctuating amplitude. Here
$Q$ is a measure of the noise intensity . For the case of B\'enard
convection at finite temperature it has been calculated in \cite{7}. Again
rolls in arbitrary directions in the $(x,y)$-plane appear with equal
weight in this distribution, in contrast to the case of the N-W-S
equation where $\bbox{k}$-vectors parallel to the $x$-axis are strongly
preferred \cite{7}.

In summary, we have given a systematic derivation
of the amplitude equation for the formation of roll patterns in
2-dimensional systems by spontaneous symmetry breaking. The derivation
is free from anisotropic scaling assumptions and establishes a
rotationally form-invariant extension of the N-W-S equation as the
fundamental amplitude equation for such systems.

\vspace{0.25cm}

\noindent
{\large\bf Acknowledgements}

\vspace{0.25cm}
This work was supported by the Deutsche Forschungsgemeinschaft through
its Sonderforschungsbereich 237 ``Unordnung und gro{\ss}e Fluktuationen''.

\end{document}